\documentclass[useAMS,usenatbib,usegraphicx]{mn2e}

\title[The central 10 pc of 3C~84]{The young radio lobe of 3C 84: inferred gas properties in the central 10 parsec}
\author[Y. Fujita et al.]{Yutaka
Fujita$^{1}$\thanks{E-mail: fujita@vega.ess.sci.osaka-u.ac.jp}, 
Nozomu Kawakatu$^{2}$, 
Isaac Shlosman$^{3,1}$, and
Hirotaka Ito$^{4}$
\\ 
$^{1}$Department of Earth and
Space Science, Graduate School of Science, Osaka University, 1-1
Machikaneyama-cho, \\ Toyonaka, Osaka 560-0043, Japan\\ 
$^{2}$Faculty of Natural Sciences, National Institute of Technology,
Kure College, 2-2-11 Agaminami, Kure, Hiroshima, 737-8506, Japan\\
$^{3}$Department of Physics and Astronomy, University of Kentucky,
Lexington, KY 40506-0055, USA\\
$^{4}$Astrophysical Big Bang Laboratory, RIKEN, Saitama 351-0198, Japan}

\begin{document}

\date{Accepted 0000 December 00. Received 0000 December 00; in original
form 0000 October 00}

\pagerange{000--000} \pubyear{0000}

\maketitle

\label{firstpage}

\begin{abstract}
 We analyse the environment of the supermassive black hole (SMBH) in the
 centre of a massive elliptical galaxy NGC~1275 in the Perseus cluster,
 hosting the radio source 3C~84. We focus on the young radio lobe
 observed inside the estimated Bondi accretion radius. We discuss the
 momentum balance between the jet associated with the lobe and the
 surrounding gas. The results are compared with the proper motion of the
 radio lobe obtained with the very long baseline interferometry. We find
 that under assumption of a high-density environment ($\ga 100\rm\:
 cm^{-3}$), the jet power must be comparable to the Eddington luminosity
 --- this is clearly inconsistent with the current moderate activity of
 3C~84, which indicates instead that the jet is expanding in a very low
 density region ($\la 1\rm\: cm^{-3}$), along the rotation axis of the
 accretion flow. The power required for the jet to expand in the
 low-density environment is comparable to the past average jet power
 estimated from the X-ray observations. We estimate the classical Bondi
 accretion rate, assuming that (1) gas accretion is spherically
 symmetric, (2) accretion is associated with the jet environment, and
 (3) the medium surrounding the jet is representative of the properties
 of the dominant accreting gas. We find that Bondi accretion is
 inconsistent with the estimated jet power. This means that either
 accretion of the cold gas in the NGC~1275 is more efficient than that
 of the hot gas, or the jets are powered by the SMBH spin.
\end{abstract}


\section{Introduction}
\label{sec:intro}

Gravitational potential energy of the accreting gas is converted into
radiation, and into kinetic and thermal energies in accretion flows onto
supermassive black holes (SMBHs). This is the essence of the active
galactic nuclei (AGN) phenomenon. Such activities include strong
outflows, sometimes associated with relativistic jets, which exert a
great impact on the surroundings \citep[e.g.][]{beg89a}.  Most of the
activities occur in the vicinity of the SMBH, which makes it difficult
to resolve.

The compact radio source 3C~84, hosted by the Seyfert galaxy NGC~1275,
may be an exceptional AGN, because it is located at the centre of the
nearby cluster (Perseus; $z=0.018$). Very long baseline interferometry
(VLBI) observations have revealed that 3C~84 exhibits complicated radio
structure on parsec scales. It consists of a bright core with a flat
spectrum and a mushroom-like southern lobe with a steep spectrum
\citep[][see Fig.~\ref{fig:asa06a}]{ven93,wal94,asa06a}. The proper
motion of the southern lobe ($\sim 0.3\rm\: mas\: yr^{-1}$) have also
been noticed in radio observations \citep{rom82,wal94}. While the
synchrotron emission from relativistic electrons provides us with
information on the detailed structure of the radio lobe and the jet
forming this lobe, not much is known about the thermal hot gas
surrounding them. This is because even with the superb resolution of
{\it Chandra}, the gas morphology within $\sim$~5~kpc from the SMBH
remains unresolved (\citealt{fab06a}, but see \citealt{won14a} for
NGC~3115). The hot gas in this region gradually accretes on to the SMBH
and fuels the AGN. In particular, the form of accretion in the immediate
vicinity of the SMBH (say within the Bondi accretion radius) can
determine the energy conversion to the radiative, kinetic, and thermal
energies \citep[e.g.][]{yua14a}. Thus, it is very important to know the
state of the hot gas there, because it determines the gas accretion on
to the SMBH.

Some radio observations have hinted at the existence of a dense gas on
scales of several parsecs. \citet{ode84a} argued about the presence of a
thermal gas of a density $\sim 2000 \rm\: cm^{-3}$ and of a temperature
$10^4$~K surrounding the core of 3C 84. This is based on an exponential
cutoff at the low-frequency end of the compact core spectrum as well as
the very low polarization at radio frequencies. Successive radio
observations showed that the absorption is spatially biased and that the
dense gas may be distributed in a form of a disc
\citep{ver94a,wal94}. This kind of dense gas has been observed in other
galactic cores \citep[e.g.][]{kam00a,pih03a}. Although detection of
absorption can constrain the properties of cold gas in the galactic
centre, such a gas probably occupies only a small volume. Faraday
rotation may also provide a clue on the gas density
\citep{pla14a}. However, it is an integrated quantity along the
line-of-sight and depends on the properties of magnetic fields.

In this paper, we explore the properties of hot gas surrounding the SMBH
in NGC~1275 using VLBI observations of the radio lobe, estimated to lie
within the Bondi radius. We focus on the fact that the evolution of the
jet associated with the lobe depends strongly on the environment as well
as on the jet power which feeds the lobe. Thus, we discuss the momentum
balance between the jet and the surrounding gas, compare the results
with the observed proper motion of the lobe, and derive information
about the environment. The jet power analysis is based on the kinematics
of the young lobe, which differs from the previous approach based on
radio power \citep{raw91a,god13}.

The redshift of the NGC~1275 ($z=0.018$) corresponds to an angular
diameter distance of 75.5~Mpc and an angular scale of $0.366\rm\: pc\:
mas^{-1}$ for the Hubble constant of $70\rm\: km\: s^{-1} Mpc^{-1}$.

\section{Models}

In this section, we explain our model for the evolution of the jets
injected by the SMBH at the centre of NGC~1275. The jet is included in
the young radio lobe (Fig.~\ref{fig:asa06a}).

\subsection{Young Radio Lobe at the Centre of NGC~1275}
\label{sec:obs}

\begin{figure}
\includegraphics[width=84mm]{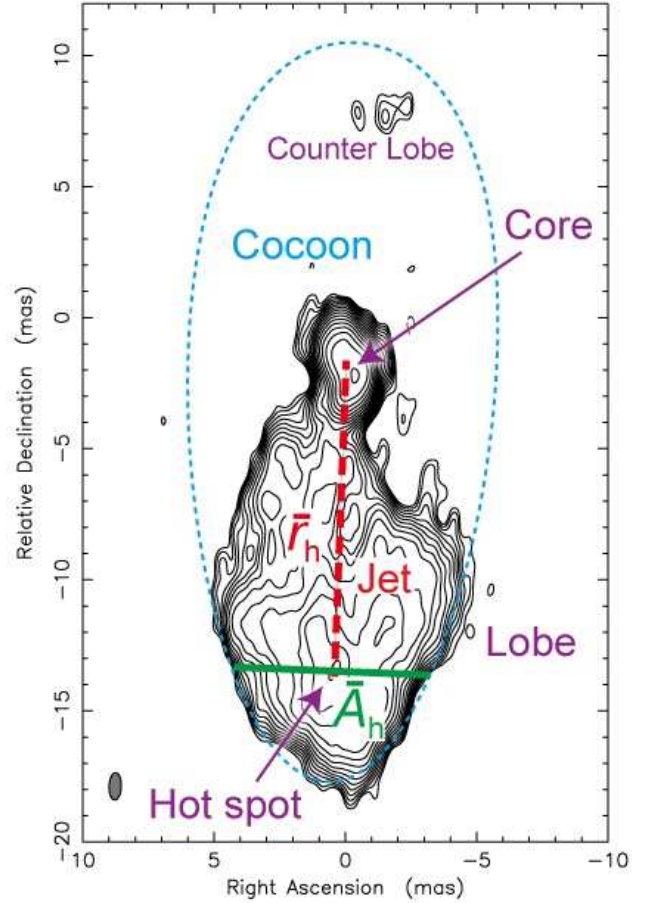} \caption{{\it VSOP} radio image of
3C~84 in 2001 (fig.~3 of \citealt{asa06a}). We evaluated the distance
from the galactic centre to the hotspot, $r_{\rm h0}$, and the
cross-section area of the cocoon head, $A_{\rm h0}$, from the lengths of
the thick dashed line and thick solid line, respectively (see text). The
thick dashed line also indicates the position of the possible jet. The
estimated cocoon containing the radio lobes is shown by the dotted
line.}  \label{fig:asa06a}
\end{figure}

First, we summarize properties of the young radio lobe associated with
3C~84 at the centre of NGC~1275; they are used as input parameters in
our models. \citet{asa06a} performed {\it VLBI Space Observatory
Programme (VSOP)} observations of 3C~84 and revealed the details of the
the radio structure. In Fig.~\ref{fig:asa06a}, the core is at the
galactic centre and the brightest spot in the southern lobe is the
hotspot at the end of the jet. The northern counter lobe is
dark. \citet{asa06a} estimated the inclination angle of the jets from
the apparent distances to the northern and southern radio lobes from the
galactic centre and found that it is $\theta = 53^\circ.1\pm
7^\circ.7$. They observed 3C~84 on~1998 August 25 at 4.916~GHz and on
2001 August~21 at 4.865~GHz. We refer to the 'current' time ($t=t_{\rm
age}$) as the time at the second observation (2001
August~21). \citet{asa06a} detected the motion of the hotspot between
the two observations and found that its apparent velocity is $v_{\rm
a0}=(0.32\pm 0.08)c$ towards the south for our Hubble constant. Thus, the
actual velocity is given by
\begin{equation}
\label{eq:va}
 \frac{v_{\rm h0}}{c} = \frac{v_{\rm a0}}
{c\sin\theta + v_{\rm a0}\cos\theta}\approx 0.32 \:,
\end{equation}
where the quantity with the subscript zero is measured at $t=t_{\rm
age}$, so $v_{\rm h0} = v_{\rm h}(t_{\rm age})$, where $v_{\rm h}(t)$ is
the velocity of the hotspot at time $t$. We expect that the radio lobes
are included in a cocoon that is filled with cosmic rays injected by the
jets. The configuration of the jet and the cocoon is shown in
Fig.~\ref{fig:asa06a} (see also fig.~1 of \citealt{kaw06a} or fig.~1 of
\citealt{ito08a}).  The velocity of the hotspot, $v_{\rm h}(t)$, is also
the advance velocity of the cocoon head.

In Fig.~\ref{fig:asa06a}, the apparent distance between the galactic
centre and the hotspot is 12.2~mas (thick dashed line). Considering the
inclination angle of the jet $\theta$, the actual distance is $r_{\rm
h0}\approx 5.6$~pc. The diameter of the radio lobe at the hotspot is
7.5~mas (thick solid line in Fig.~\ref{fig:asa06a}). Thus, the
cross-section area of cocoon head is $A_{\rm h0}=\pi(7.5/2)^2 {\rm\:
mas^2} =5.9\rm\: pc^2$.

We note here that there are some uncertainties in the radio
observations. \citet{asa06a} estimated the expansion velocity, $v_{\rm
a0}=0.32\: c$ from the apparent motion of the peak of the lobe surface
brightness, by measuring the distance from the core. Although the core
has two components, \citet{asa06a} used the position of the resolved
most northern component as the reference point (see their fig.~3). Thus,
the ambiguity of the core position is expected to be small. On the other
hand, the motion of the peak may reflect the pattern velocity of the
lobe, and not the physical expansion velocity. Moreover, the {\it
instantaneous} hotspot may not be moving with the working surface
represented by $A_{\rm h}$. If there is an error of a factor of 2 in the
apparent velocity of the hotspot ($v_{\rm a0}$), which is probably too
large, it will cause an error of a factor of 3 or 4 in the jet power
(equations~\ref{eq:va} and~\ref{eq:Lj} below). Lastly, \citet{asa06a}
obtained the inclination angle of the jet, $\theta = 53.1^\circ\pm
7.7^\circ$, from the ratio of the distances from the galaxy centre to
the south and north lobes. However, it has been indicated that
free--free absorption can affect the northern lobe
\citep{wal03a}. Hence, if the absorption is highly inhomogeneous, it may
affect the estimated distances to the lobes. We also note that
\citet{lis09a} took a different approach, by observing the apparent
motions of both southern and northern lobes by fitting them with a
Gaussian model, and by determining the inclination angle of the jets to
be $\theta \sim 11^\circ$ from the velocities of the southern lobe
($\sim 0.31\: c$) and the northern lobe ($\sim 0.08\: c$). The former is
consistent with \citet{asa06a}. The Gaussian fitting may be less
sensitive to the pattern change in the lobes, although the complex core
component may still have an influence on the measurement \citep{lis09a}.
We have confirmed that our conclusions about the jet power is almost the
same (within a factor of a few), even if we adopt $\theta = 11^\circ$
instead of $53.1^\circ\pm 7.7^\circ$. However, this value $\theta$ has a
problem with the age of the jet (see Section~\ref{sec:result}). We also
note that recent {\it Fermi} observations have shown that the
inclination angle of the jets should be $\theta \sim
25^\circ$--$32^\circ$ from the gamma-ray spectrum
\citep{abd09b}. Considering the overall insensitivity of the results to
the inclination angle, we adopt $\theta=53^\circ.1$ from now on.

\subsection{Ambient Gas Profile}
\label{sec:n}

\begin{figure}
\includegraphics[width=84mm]{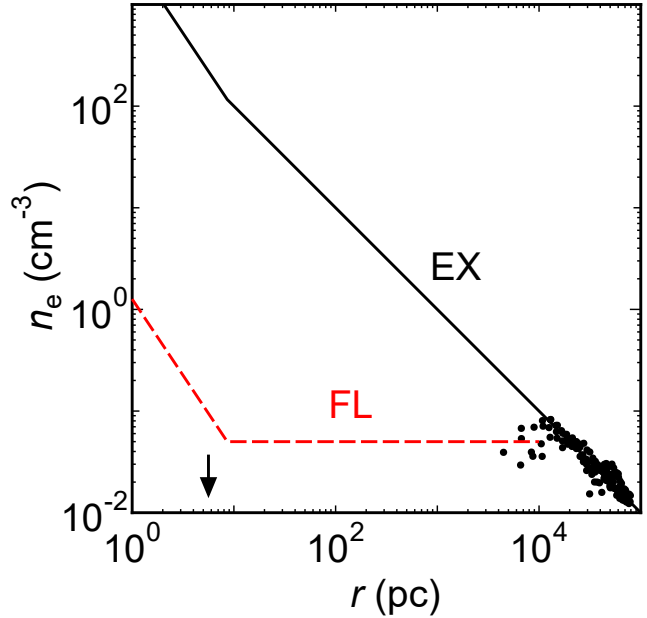} \caption{Assumed density and
temperature profiles. Dots are the {\it Chandra} observations
\citep{fab06a}. The arrow shows the length of the jet ($r_{\rm
h0}$).}  \label{fig:n}
\end{figure}

In our discussion of the evolution of the jet, including the radio lobe,
we need to specify the properties of the surrounding gas. {\it Chandra}
observations have shown that the electron number density of the hot gas
increases and the temperature decreases towards the galactic centre, and
they reach $n_{\rm e}\sim 0.1\rm\: cm^{-3}$ and $T\sim 3$~keV,
respectively, at $r\sim 10$~kpc from the centre \citep[e.g.][see
Fig.~\ref{fig:n}]{fab06a,raf06a,gra08a}. At $r\sim 5$--10~kpc, the hot
gas is strongly affected by the X-ray cavities created by past AGN
activities, and the density is $n_{\rm e}\sim 0.05\rm\: cm^{-3}$, and
the temperature is $T\sim 3$~keV \citep[][see
Fig.~\ref{fig:n}]{fab06a}. There is no information on the gas for $r\la
5$~kpc.

Assuming that the gas temperature around the galactic centre ($r\la
10$~kpc) is $T=3$~keV and the mass of the central black hole is
$M_\bullet=8\times 10^8\: M_\odot$ \citep{sch13a}, the Bondi radius is
\begin{equation}
\label{eq:rB}
 r_{\rm B}=\frac{2 G M_\bullet}{c_{\rm s}^2}=8.6\rm\: pc\:,
\end{equation}
where $c_{\rm s}$ is the sound velocity. This radius is comparable to
the current length of the jet ($r_{\rm h0}\sim 5.6$~pc,
Section~\ref{sec:obs}). Since $r_{\rm B}$ is much smaller than the
innermost radius for the X-ray observations ($r\sim 5$--10~kpc), we
extrapolate the observed gas density profiles towards $r_{\rm B}$ using a
power law:
\begin{equation}
 n_{\rm e}(r) = n_{\rm e,obs}\left(\frac{r}{r_{\rm obs}}\right)^{-b}\:,
\end{equation}
for $r_{\rm B}<r<r_{\rm obs}=10$~kpc. We assume that the gas
distribution is spherically symmetric for simplicity. The density
$n_{\rm e,obs}$ and the index $b$ is the parameters for which we
consider two extreme cases. In the first model (model~EX), we just
extrapolate the density profile for $r\ga 10$~kpc (Fig.~\ref{fig:n}),
and we set $n_{\rm e,obs}=0.1\rm\: cm^{-3}$ and $b=1$. In this case, the
gas density at $r_{\rm B}$ is $n_{\rm e,B}=116\rm\; cm^{-3}$. In the
second model (model~FL), we assume that a flat density profile, and we
set $n_{\rm e,obs}=n_{\rm e,B}=0.05\rm\: cm^{-3}$ and $b=0$
(Fig.~\ref{fig:n}). This may reflect a case where thermal instability of
the gas is effective and most of the hot gas turns into cold gas
\citep{bar12a,gas12c,mcc12a,sha12a,guo14a,mee15a}. Since simulations
show $b>0$ \citep[e.g.][]{gas13a}, we expect that $b=0$ is a lower
limit.  While the remaining hot gas occupies most of the volume, the
density is significantly reduced around the centre.

We assume that the density profile for $r<r_{\rm B}$ is given by another
power law:
\begin{equation}
\label{eq:neB}
 n_{\rm e}(r) = n_{\rm e,B}\left(\frac{r}{r_{\rm B}}\right)^{-\alpha}\:,
\end{equation}
where $n_{\rm e,B}=n_{\rm e}(r_{\rm B})$, and $\alpha$ is the parameter.
In this region, the gas distribution should be strongly affected by the
SMBH. If the gas falls nearly freely into the SMBH, the infall velocity
is $v\propto r^{-1/2}$ from the energy conservation, and $\alpha$ should
be 1.5, from the mass conservation. This profile approximates the one
for the classical Bondi accretion \citep{bon52a}.  On the other hand,
$0.5\leq\alpha\leq 1.5$ was suggested by \citet{bla99a} in the context
of advection-dominated inflow-outflow solutions (ADIOS). Moreover,
recent numerical simulations of a gravitational collapse in the presence
of the gas cooling have shown that the slope should be $0\leq\alpha\leq
1.5$ \citep[e.g.,][]{cho13a,cho15a}. We have confirmed that the results
are not sensitive to the value of $\alpha$ as long as $0\la\alpha\la
1.5$. Thus, we assume $\alpha=1.5$ from now on.

\begin{table*}
\centering
 \begin{minipage}{160mm}
  \caption{Results}\label{tab:result}
  \begin{tabular}{@{}lccccccccc@{}}
   \hline
   Model & $n_{\rm e}(r_B)$ & $\alpha$ 
& $n_{\rm e}(r_{\rm h0})$ & $t_{\rm age}$ & $2L_{\rm j}$
& $2L_{\rm j}/L_{\rm Edd}$   & $2L_{\rm j}/P_{\rm cav}$
& $P_{\rm B}$ & $2L_{\rm j}/P_{\rm B}$\\
        & $(\rm cm^{-3})$ &
        & $(\rm cm^{-3})$ & (yr)  & $(\rm erg\: s^{-1})$
     &  &   & $(\rm erg\: s^{-1})$    & \\
  \hline
EX  & 116 & 1.5 & 222 & 46 & $1.3\times 10^{47}$
   & 1.3 & 869 & $9.8\times 10^{44}$ & 133 \\
FL  & 0.05 & 1.5 & 0.096 & 46 & $5.6\times 10^{43}$
   & $5.6\times 10^{-4}$ & 0.37 & $4.2\times 10^{41}$ & 133 \\
  \hline 
\end{tabular}

 Note --- Column (1): Model name. Column (2) Gas density at the Bondi
 radius. Column (3): Index of gas profile within the Bondi
 radius. Column (4): Gas density at the tip of the jet. Column (5): Jet
 age. Column (6): Jet power. Column (7): Ratio of jet power to Eddington
 luminosity. Column (8): Ratio of jet power to cavity power. Column (9)
 Bondi accretion power. Column (10): Ratio to jet power to Bondi
 accretion power.
\end{minipage}
\end{table*}

\subsection{Evolution of the Jet}
\label{sec:jet}

The momentum balance along the jet is given by
\begin{equation}
\label{eq:mom}
 \frac{L_{\rm j}}{v_{\rm j}} = \rho(r_{\rm h})
v_{\rm h}(t)^2 A_{\rm h}(t)\:,
\end{equation}
where $L_{\rm j}$ is the kinematic power of the jet, $v_{\rm j}$ is the
velocity of the jet, $r_{\rm h}$ is the distance to the hotspot from
the galactic centre, and $A_{\rm h}$ is the cross-section area of the
cocoon head. We assume that $L_{\rm j}$ is time-independent and $v_{\rm
j}=c$.  We also assume that the opening angle of the cocoon head,
$\theta_{\rm h} = \arctan(A_{\rm h}^{1/2}/(\pi^{1/2} r_{\rm h}))$, is
constant in time. Since the jet lies within the Bondi radius at present
($r_{\rm h0}<r_{\rm B}$), the density profile of the ambient gas is
given by equation~(\ref{eq:neB}) and can be rewritten as
\begin{equation}
\label{eq:rhoa}
 \rho(r)=\rho_{\rm B} (r/r_{\rm B})^{-\alpha} \:.
\end{equation}
From equations~(\ref{eq:mom}) and~(\ref{eq:rhoa}), we obtain
\begin{equation}
\label{eq:Lj}
L_{\rm j} = \rho(r_{\rm h0})c v_{\rm h0}^2 A_{\rm h0}\:,
\end{equation}
and
\begin{equation}
\label{eq:tage}
 t_{\rm age} = \frac{2 r_{\rm h0}}{4-\alpha}
\left(\frac{r_{\rm h0}}{r_{\rm B}}\right)^{-\alpha/2}
\left(\frac{L_{\rm j}}{\rho_{\rm B} c
				A_{\rm h0}}\right)^{-1/2} \:.
\end{equation}

\section{Results}
\label{sec:result}

We take $r_{\rm h0}=5.6$~pc, $A_{\rm h0}=5.9\rm\: pc^2$, and $v_{\rm
h0}=0.32c$ (Section~\ref{sec:obs}).  From equations~(\ref{eq:neB}),
(\ref{eq:Lj}), and (\ref{eq:tage}), we obtain $n_{\rm e}(r_{\rm h0})$,
$L_{\rm j}$ and $t_{\rm age}$, respectively. Our results are summarized
in Table~\ref{tab:result}.  Since there are two jets (north and south),
the total jet power is represented by $2L_{\rm j}$. The power decreases
as the ambient gas density at $r=r_{\rm h0}$ or $n_{\rm e}(r_{\rm h0})$
decreases, and thus model~EX gives the higher power
(Table~\ref{tab:result}). This is because a larger density of the
surrounding gas requires a larger jet power to excavate for a given
current advance speed of the jet
(equations~\ref{eq:rhoa}--\ref{eq:tage}).

The age of the jet $t_{\rm age}$ does not depend on $n_{\rm e}(r_{\rm
h0})$ for a given $\alpha$ and the fixed $v_{\rm h0}$. Again, this is
because a larger ambient density requires a larger a jet power to
excavate; $L_{\rm j}/\rho_{\rm B}$ does not depend on $\rho_{\rm B}$ in
equation (\ref{eq:tage}) (see also equations~\ref{eq:rhoa}
and~\ref{eq:Lj}). Since $t=t_{\rm age}$ corresponds to 2001, the jet was
born in 1955. Observations have indicated that an outburst started in
1959 \citep[e.g.][]{nes95a}. Our model is almost consistent with the
idea that this outburst corresponds to the formation of the radio lobes.
We note that regardless of the density profile (EX or FL), models with
the inclination angle of $\theta=11^\circ$ (Section~\ref{sec:obs})
predict that the jet was born before 1940, which is not consistent with
the observed outburst in 1959.

Next, we compare the obtained jet power with other quantities associated
with the SMBH and AGN in this galaxy. The Eddington luminosity is given
by
\begin{equation}
 L_{\rm Edd} = 1.26\times 10^{38}\left(\frac{M_\bullet}{M_\odot}\right)
\rm\: erg\: s^{-1}\:.
\end{equation}
Since gas kinematics suggests $M_\bullet=8\times 10^8\: M_\odot$
\citep{sch13a} for NGC~1275, the luminosity is $L_{\rm Edd}=1.0\times
10^{47}\rm\: erg\: s^{-1}$.  In Table~\ref{tab:result}, we present
$2L_{\rm j}/L_{\rm Edd}$. Model~EX predicts that the ratio is close to
one. However, such a high ratio is accepted only for extremely active
AGNs, which radiate efficiently as QSOs. It is not likely to be realized
for moderate AGNs, including 3C~84 \citep[e.g.][]{yua14a}. In fact, the
X-ray luminosity of 3C~84 is $\sim 10^{43}\rm\: erg\: s^{-1}$, much
smaller than $L_{\rm Edd}$ \citep{don04b}.

Another interesting quantity is the cavity power $P_{\rm cav}$, which is
required to inflate X-ray cavities observed around an AGN. The cavities
are thought to be the relics of cocoons. In the case of NGC~1275,
cavities are located at $r\ga 1$~kpc from the centre \citep{fab06a}, and
the volume $V_{\rm cav}$ can be estimated from their sizes. The total
energy required to create a cavity is equal to its enthalpy and it is
given by
\begin{equation}
 E_{\rm cav}=\frac{\gamma_{\rm c}}{\gamma_{\rm c}-1}
p_{\rm gas}V_{\rm cav}\:,
\end{equation}
where $\gamma_{\rm c}=4/3$ is the adiabatic index for relativistic gas,
and $p_{\rm gas}$ is the pressure of the gas surrounding the cavities,
which can be estimated from X-ray observations. The cavity power is
defined by

\begin{equation}
 P_{\rm cav} = E_{\rm cav}/t_{\rm burst}\:,
\end{equation}
where $t_{\rm burst}$ is the average time between outbursts of the AGN.
\citet{raf06a} estimated that the cavity power with NGC~1275 is $P_{\rm
cav}\sim 1.5\times 10^{44}\rm\: erg\: s^{-1}$, which may represent the
average jet power over the period of $t_{\rm burst}\sim 10^7$~yr. In
Table~\ref{tab:result}, we provide the ratio $2L_{\rm j}/P_{\rm
cav}$. For model~EX, this ratio is much larger than unity, which means
that the current jet power exceeds substantially the past average. Since
the probability to observe the jet activity is $\sim (2L_{\rm j}/P_{\rm
cav})^{-1}$, it is unlikely that such activity is actually occurring at
present. On the other hand, the same ratio in model~FL is close to unity
(Table~\ref{tab:result}), which means that the current jet activity is
not very different from the past average. Note, that the energy inside
the cavity or the cocoon could also be estimated by means of the
minimum-energy condition. Using this condition, \citet{nag09a} estimated
that the jet power is $\sim 10^{42}\rm\: erg\: s^{-1}$, which is more
than one orders of magnitude smaller than the predictions of model FL
(Table~\ref{tab:result}). However, it has been indicated that this
condition is not met for NGC~1275, and that the cavities may contain
hidden relativistic particles, such as protons \citep{fab02a,nag09a}.

The Bondi accretion rate is given by
\begin{equation}
\label{eq:dotMB}
 \dot{M}_{\rm B} = (\pi/4) c_{\rm s,0}\rho_{\rm B} r_{\rm B}^2\:,
\end{equation}
\citep{bon52a}. Assuming efficiency of 10\%, the maximal power
released from the vicinity of the SMBH by the Bondi accretion
is
\begin{equation}
 P_{\rm B} = 0.1 \dot{M}_{\rm B} c^2\:.
\end{equation}
For reference reasons only, we have calculated the ratio $2L_{\rm
j}/P_{\rm B}$ for all the models, using the idealized spherical Bondi
accretion rate.

The ratio $2L_{\rm j}/P_{\rm B}$ depends on the accreting gas
properties, i.e., on its density profile, and its temperature.  Thus,
models EX and FL, for a given $\alpha$, result in the same values of
$2L_{\rm j}/P_{\rm B} \gg 1$. This means that the characteristic Bondi
accretion rate cannot provide enough power to sustain the observed jets
regardless of the gas properties around the Bondi radius. We emphasize
that this argument can be applied when the accretion is spherically
symmetric, and the jet environment and accretion are correlated.

We have assumed that the jet is interacting with the hot gas in
NGC~1275. However, the central region of the galaxy is very complicated
and the gas may be multiphase. For example, the jet may be interacting
with dense cool gas $\ga 1000\rm\: cm^{-3}$ mentioned in
Section~\ref{sec:intro}. In this case, however, the jet power must be
even larger than that for model~EX (Table~\ref{tab:result}) in order to
be consistent with the current jet velocity. Such an extreme jet power
($2L_{\rm j}/L_{\rm Edd}\gg 1$) is very unlikely. On the other hand, the
lobe may be expanding into an existing larger scale radio lobe, in which
gas density is much smaller than that in model FL. If this is the case,
the jet power and $2L_{\rm j}/P_{\rm cav}$ are also much smaller than
those in model FL (Table~\ref{tab:result}). This means that the current
jet activity is much weaker than the time average.

In summary, the ratios $2L_{\rm j}/L_{\rm Edd}$ and $2L_{\rm j}/P_{\rm
cav}$ show that the lower density model (FL) is preferable to the higher
density model (EX) in terms of the current and past moderate activities
of the AGN. The classical Bondi accretion model can be rejected, because
it cannot provide enough jet power regardless of the ambient gas
density.

\section{Discussion and Conclusions}
\label{sec:diss}

The results in Section~\ref{sec:result} indicate that the gas density
around the cocoon is likely to be $\la 1\rm\: cm^{-3}$ (model~FL). This
low density may result from thermal instabilities, as we mentioned in
Section~\ref{sec:n}. Another possibility is that the hot gas is
distributed highly anisotropically, and its density in the direction of
the jets is substantially smaller than in other directions, perhaps
because previous generations of jets have drilled tunnels. The jet which
has been analysed here is expanding nearly freely through low-density
environment (model~FL) at present. Since this reduces energy loss of the
jet during the expansion against the ambient gas, it may contribute to
the efficient energy transport from the jets to the intracluster medium,
and contribute to the suppression of a massive cooling flow
\citep[e.g.][]{fuj13a}.

On the other hand, some radio observations have detected absorbing
material with a density of $>1000\rm\: cm^{-3}$ on the scale of several
parsecs \citep{ode84a}. A possible solution is that the dense gas forms a
cold disc around the SMBH, and the jets are launched perpendicular to
the disc. \citet{wal03a} indicated the northern counter lobe shown in
Fig.~\ref{fig:asa06a} is absorbed. This suggests that the size of the
disc is at least comparable to the apparent distance to the northern
lobe ($\sim 3$~pc).

Our results show that the classical Bondi accretion cannot provide
enough power to sustain the observed jets. Addition of angular momentum
to the accreting gas will not improve the situation in any substantial
way, because while the gas density decreases along the the rotation axis
(which is the direction of jet propagation), the gas accretion rate to
the SMBH decreases as well. The gas will be diverted away from the
rotation axis towards the equatorial plane and may accumulate there
\citep[e.g.][]{pro03a,kru05a}. The fate of this gas is outside the scope
of this paper, but it can be either converted to stars, expelled in a
wind, or be accreted by the SMBH in the disc accretion mode.

This indicates that it is the cold gas, rather than the hot gas, that
fuels the SMBH \citep{shl89a,piz10a,gas13a}. However, such process
requires an efficient mechanism which removes the angular momentum from
the gas, such as gravitational \citep{shl89a,shl90a} or magnetic
\citep{bland82a,emmer92a} torques. The cold gas can originate in thermal
instabilities, and can form a disc around the SMBH. If the disc is
turbulent, this would decrease the mass inflow time-scale and
correspondingly increase the accretion rate on to the SMBH
\citep{shl90a,kaw08b}. Alternatively, the jets can be powered by the
SMBH spin \citep{bla77}. The existence of a cold disc gas can be
confirmed by future observations (e.g. ALMA).

In this model, we do not consider relativistic effects in the jet model
(Section~\ref{sec:jet}). However, they do not significantly affect the
results because of the short duration of the relativistic expansion of
the jet ($v_{\rm h}\sim c$). Using equations~(\ref{eq:mom})
and~(\ref{eq:rhoa}) and the assumption of a constant opening angle, one
can obtain $v_{\rm h}\propto t^{(\alpha-2)/(4-\alpha)}$. Thus, the jet
expands as $v_{\rm h}\propto t^{-0.2}$ for $\alpha = 1.5$. Therefore,
the duration of the relativistic expansion is only $\sim 0.003\: t_{\rm
age}$ for a given $v_{\rm h0}=0.32\: c$. In fact, the jet evolution
(e.g. $r_{\rm h}(t)$) is insensitive to the choice of the start time of
the expansion \citep{ito08a}.

To summarize, we have studied the environment of the young radio lobe
observed at $\la 10$~pc from the SMBH at the centre of the galaxy
NGC~1275 (or the radio source 3C~84) in the Perseus cluster. The lobe is
located inside the estimated Bondi accretion radius. The lobe should be
associated with the cocoon created by jet activities in the AGN, and the
evolution of the jet should depend on the jet power and the properties
of the hot ambient gas.

Using a simple model and comparing it with the proper motion of the lobe
obtained through VLBI observations, we constrain the jet power and the
environment in the vicinity of the SMBH. We find that a low-density
environment $\la 1\rm\: cm^{-3}$ is more plausible, considering the
current and past modest activities of the AGN. This low density hot gas
can result from the thermal instabilities in a denser precursor
gas. Since the existence of a high-density gas ($> 1000\rm\: cm^{-3}$;
\citealt{ode84a,ver94a,wal94}) has been indicated in this region, we
conjecture that the gas is highly inhomogeneous and it is plausibly
forming a disc around the SMBH. Assuming that the accretion is
spherically symmetric and is correlated with the gas properties around
the jet, and assuming that the gas close to the jet has the properties
of the dominant accreting gas, we show that the classical Bondi
accretion rate is not consistent with our predicted jet power, and,
therefore, can be rejected. The presence of angular momentum in the
accreting gas will not improve the situation. A more realistic scenario
with the seed angular momentum in the accretion flow makes it plausible
that the gas will accumulate in the equatorial plane of the flow, where
gravitational or magnetic torques trigger an efficient loss of the
angular momentum in the cold gas. This means that either accretion of
the cold gas dominates over that of the hot gas, or the jets are powered
by the SMBH spin.

\section*{Acknowledgments}

We thank the anonymous referee for useful comments. We also thank
K.~Asada for useful discussion and providing us the image of 3C~84. This
work was supported by International Joint Research Promotion Program and
Challenge Support Program by Osaka University. This work was also
supported by KAKENHI (Y.~F.: 15K05080). N.~K. acknowledges the financial
support of Grant-in-Aid for Young Scientists
(B:25800099). I.~S. acknowledges partial support from the NSF and
STScI. STScI is operated by AURA, Inc., under NASA contract NAS
5-26555. H.~I. acknowledges the financial support of Grant-in-Aid for
Young Scientists (B:26800159).

\end{document}